\documentstyle[12pt,fleqn]{article}
\def\la{\langle}
\def\ra{\rangle}
\def\beeq{\begin{equation}}
\def\eneq{\end{equation}}
\def\beeqa{\begin{eqnarray}}
\def\eneqa{\end{eqnarray}}

\addtocounter{section}{-1}
\begin{document}

\begin{center}

\vspace{2cm}

{\large {\bf {Optical excitations in diphenylacetylene based dendrimers\\
studied by a coupled exciton model with off-diagonal disorder
} } }

\vspace{1cm}

{\rm Kikuo Harigaya\footnote[1]{E-mail address: 
\verb+harigaya@etl.go.jp+; URL: 
\verb+http://www.etl.go.jp/+\~{}\verb+harigaya/+}}

\vspace{1cm}

{\sl Physical Science Division,
Electrotechnical Laboratory,\\ 
Umezono 1-1-4, Tsukuba 305-8568, 
Japan}\footnote[2]{Corresponding address}\\
{\sl National Institute of Materials and Chemical Research,\\ 
Higashi 1-1, Tsukuba 305-8565, Japan}\\
{\sl Kanazawa Institute of Technology,\\
Ohgigaoka 7-1, Nonoichi 921-8501, Japan}

\vspace{1cm}

(Received~~~~~~~~~~~~~~~~~~~~~~~~~~~~~~~~~~~)
\end{center}

\vspace{1cm}

\noindent
{\bf Abstract}\\
A phenomenological coupled exciton model is proposed in order 
to characterize optical excitations in extended dendrimers.  
An onsite exciton state is assigned at each phenyl rings 
and a nearest neighbor hopping integral which obeys the 
Gaussian distribution is considered between the exciton states.  
The decreasing optical excitation energy with respect to 
the dendrimer size indicates the presence of exciton funnels 
along the backbone of the dendrimers.  Therefore, the extended 
dendrimers can work as artificial fractal antenna systems 
which capture energy of light.  Dynamics of an exciton is 
also investigated by solving time development of a wavefunction
of optical excitations.   It is shown that a damping term with
a certain magnitude is necessary in order that optical
excitations captured at the outer edge of the supermolecule
move to the central areas of the molecule.

\vspace{1cm}

\pagebreak

\section{Introduction}

Recently, the dendrimer supermolecules with antenna structures
have been investigated extensively.  After capture of light
at the outer edges of the molecule, generated excitons
migrate along the legs of the molecular structures and carry
energy obtained from light.  Then, the excitons move to a certain 
core to localize there, or move to the center of the supermolecule
in order to emit light by recombination of electrons and holes.
Because energy of light is captured and the energy is transfered
by excitons, the supermolecules might act as artificial
molecular antenna.  Therefore, the design of the molecular
structures and their optical properties are quite attractive
in view of scientific interests as well as their potential for
technological applications.

One example of antenna supermolecules has dendrimeric 
structures.  It is a family of molecules composed of 
phenyl rings and acetylene units, namely, extended dendrimers 
[1-5].  Their geometrical structures are illustrated in Fig. 1.
Figure 1 (a) shows the diphenylacetylene with
its abbreviated notation.  Figure 1 (b) shows the family of
extended dendrimers: D4, D10, D25, D58, and D127.  The number
in their names means the number of phenyl rings in the molecules.
In the D4 and D10 dendrimers, each leg is composed of two 
single bonds and one triple bond.  In D25, three central 
legs are made of one phenyl rings and two short legs (shown as 
Fig. 1 (a)).  In D58, three central legs are composed of two 
phenyl rings and three short legs (Fig. 1 (a)).  The next 
connecting legs via the phenyl vertex are like the central legs 
of D25.  In D127, the three central legs are made of three phenyls
and four short legs (Fig. 1 (a)).  In this way, the extended
dendrimers have fractal molecular structures, and the length
of the central legs becomes longer as the size of the dendrimer
becomes larger.

Optical experiments of the extended dendimers [3-5]
show that the optical gap decreases as the dendrimer 
size becomes larger.  The optical spectra have features
which can be understood as contributions from legs
of phenylacetylene oligomers.  The energy of each
feature agrees with that of oligomers.  The excitation
energy becomes smaller as the length of the legs
becomes larger.  Therefore, the presence of exciton
migration pathways along the legs or the backbone of the
supermolecules has been concluded.

Another example of the antenna supermolecules has 
structures with four zinc-containing tetraarylporphylins
linked to a central, unmetallated porphylin through ethyne
bonds [6].  They show migration characters of excitons,
and are interesting as light harvesting antennas.
Such alternative molecular designs, replacement
of molecules at the vertexes, different structure
of leg chains, and so on, have been investigated intensively.
Although various molecular structures are of interests,
we would like to concentrate upon optical excitation
properties of the extended dendrimers in this paper
because of the varieties of possible structures which
will require much more theoretical efforts in future.

Theory of optical excitations in the dendrimers has not
been reported so often, and for example energy transfer 
has been investigated by solving phenomenological 
probability process equations [7].  The rates of the
excitation flow along the legs of the dendrimer have
been introduced, and the mean passage time has been
obtained theoretically.

The first purpose of this paper is to propose a phenomenological 
coupled exciton model and characterize optical excitations 
in extended dendrimers.  By sample average over sets of 
off-diagonal disorder, we will obtain a decreasing optical 
excitation energy with respect to the dendrimer size, and
show exciton funnels are present along the backbone of the 
dendrimers.  The second purpose is to simulate flows of 
excitons by looking at time development of a wavefunction
of optical excitations.   We will show an importance of a 
damping term with a certain magnitude in order that optical
excitations captured at the outer edge of the supermolecule
move to the central areas of the molecule.  We note that
a part of this paper -- characterization of the parameters
for excitons -- has been reported in the short communication [8].

This paper is organized as follows.  In Sec. II, the coupled
exciton model with off-diagonal disorder is explained.  
Section III is devoted to characterization of optical 
excitations by comparison of the theory with experiments.
In Secs. IV and V, we study dynamics of exciton flow in
the supermolecules.  Equilibrium and nonequilibrium dynamcses
are studied in Secs. IV and V, respectively.  The paper
is summarized in Sec. VI.

\section{Formalism}

In this section, we give rise to a new theoretical model 
composed of coupled exciton states with off-diagonal disorder.
We note that the theoretical model of dipole moments [9] 
has been used in order to characterize optical excitations 
of the photosynthetic unit of purple bacteria, which is the 
biological analog of the dendrimeric supermolecule.  In the present 
model, an onsite exciton state is assigned at each phenyl ring.  
There are two possibilities for the nearest neighbor interactions:
(1) When the interactions occur by dipole-dipole couplings,
the direction of the transition dipole moment is by
no means parallel with the electric field of light and
cannot be spatially correlated.  Interaction strengths 
between neighboring dipole moments may vary among positions 
of dipole pairs.  They can be looked as randomly distributed.
Therefore, we assume that the nearest neighbor interactions
obey the Gaussian distribution function.  (2) The second candidate
of the interaction between neighboring exciton states is
an exciton flow characterized with the strengths 
$t_e t_h / \Delta_{\rm ex}$ by perturbation, where $t_e$ and 
$t_h$ are hopping integrals of electrons and holes, 
$\Delta_{\rm ex}$ is the excitation energy of the electron 
hole pair.  We assume that the mean value of the interaction 
is zero and the standard deviation of the interaction $J$ is 
one of theoretical parameters in the Gaussian distribution.
Another theoretical parameter in the model is the site 
energy $E$ which specifies the central energy position of 
excitons in the optical spectra.  The following is our 
tight binding model:
\beeq
H = E \sum_i | i \ra \la i | 
+ \sum_{\la i,j \ra} J_{i,j} ( | i \ra \la j | + {\rm h.c.} ),
\eneq
where $i$ means the $i$th site of the phenyl ring, 
$| i \ra$ is an exciton state at the site $i$,
the sum with $\la i,j \ra$ is taken over neighboring
pairs of sites, and the distribution of $J_{i,j}$ is 
determined by the Gaussian function,
\beeq
P(J_{i,j})=\frac{1}{\sqrt{4\pi}J}
{\rm exp}[-\frac{1}{2} (\frac{J_{i,j}}{J})^2].  
\eneq
Here, the disorder effects appear in the off-diagonal
matrix elements of the hamiltonian.  Such the disorder
model is named as ``off-diagonal".  When the disorder
effects appear in the diagonal matrix elements, the
terminology ``diagonal disorder" is used.  In other words,
the off-diagonal disorder model which modulates couplings
of bonds is called as  ``bond disorder", and the diagonal 
disorder model which perturbs site energies is named as 
``site disorder".
Though detailed theoretical treatments are not the same 
altogether, the related exciton models have been used for 
the photosynthetic unit of purple bacteria [9] and the 
J-aggregate systems [10-12].  The diagonalization of eq. (1) 
gives energies of one exciton states measured from the 
energy of the ground state.

\section{Optical absorption}

The model eq. (1) is diagonalized numerically for the five 
types of the extended dendrimers: D4, D10, D25, D58, and D127.
The lowest eigenvalue always gives the energy position
of the optical absorption edge because the state
with the lowest energy is always allowed for dipole 
transition from the ground state.   This is checked
by looking at the parity of the wave function for each
dendrimer.   In TABLE I, we show the energy of the
absorption edge as a function of the parameters $E$ and $J$.
Here, the number of disorder samples is 10000, and this 
gives the well converged average value of the optical
excitation energy.

Figure 2 shows one example of the comparison of the 
calculation with experiments for the parameters 
$E = 37200{\rm cm}^{-1}$ and $J = 3552{\rm cm}^{-1}$.  
We find fairly good agreement between the experiments and 
calculations.  Two results have the trend that the lowest 
optical excitation energy decreases as the dendrimer 
size becomes larger.  Therefore, it is clarified that the 
presence of exciton migration funnels is well described 
by the present coupled exciton model with off-diagonal 
disorder.  The interaction strength $J$ is one order of 
magnitudes larger than that of the purple bacteria [9], 
indicating the stronger contact between neighboring exciton
states.  When the flow of excitons occur by hopping of 
electrons and holes, the interaction strength is characterized
as $t_e t_h / \Delta_{\rm ex}$ by perturbation.  With assuming
$t_e \sim t_h \sim 0.5t$ and $\Delta_{\rm ex} \sim 2t$ where $t$ 
is the resonance integral of the $\pi$ orbitals of the
phenyl ring, we obtain a characteristic magnitude:
$|J| \sim 0.1 t$.  This quantity is also of the same
order of magnitudes with the above parameter $J = 3552{\rm cm}^{-1}$,
because $t \sim 2{\rm eV} \sim 25000{\rm cm}^{-1}$.
Thus, our theoretical parameter can characterize exciton
flows along the legs of dendrimers very well.

The extended dendrimers are the rare example where 
$\pi$-conjugated electron systems are present along
the acetylene based legs.  In most of dendrimers 
(see the recent review [13] for example), the systems 
are composed of $\sigma$-bondings rather than $\pi$-bonds.
Owing to the presence of exciton funnels composed of 
$\pi$-bonds, the extended dendrimers can work as  
an artificial fractal antenna which captures
energy of light.

In the compact dendrimers (the another form of 
diphenylacetylene based dendrimers), the optical absorption 
edge less depends on the molecule size [3].  This might come 
from the very huge steric repulsions among neighboring legs, 
and therefore the mutual interactions between legs are hindered 
easily by geometric effects.  The nearly constant absorption 
edge with respect to the dendrimer size indicates that values 
of the interaction strengths $J_{i,j}$ of the compact dendrimers 
are much smaller than those of the extended dendimers when 
the present coupled exciton model is applied to.  Such the 
difference of the parameter values can characterize the 
different dependence on the system size of the extended 
and compact dendimers.

\section{Equilibrium dynamics}

After the characterization of the optical absorption spectra
in the previous section, we shall study exciton propagations 
in dendrimers by looking at time development of wavefunctions 
in the following two sections.  We will focus on the largest 
dendrimer D127, hereafter.  This supermolecule is of most 
interests among the five dendrimers.

\subsection{Exact time development}

The dynamics of an exciton which is located at the site 1
at the time $t=0$ (see Fig. 1 (b) for the site numbers 
in D127) can be solved exactly.  When the eigenstates 
$| \mu \ra$ and their energies $\omega_\mu$ for a disorder set 
are obtained, the initial wavefunction $|\Psi_0 \ra $ of the 
exciton is expanded as:
\beeq
|\Psi_0 \ra = \sum_\mu a_\mu | \mu \ra.
\eneq
At the time $t$, this wavefunction becomes
\beeqa
|\Psi \ra &=& e^{-iHt} |\Psi_0 \ra \\
&=& \sum_{\mu,i} a_\mu b_{\mu,i} e^{-i \omega_\mu t} |i \ra,
\eneqa
where $b_{\mu,i}$ is an expansion coefficient in the
site representation, $|\mu \ra = \sum_i b_{\mu,i} | i \ra$.
The probability amplitude at the site $i$ is calculated as
\beeq
| \sum_\mu e^{i \omega_\mu t} a_\mu b_{\mu,i} | ^2 
= \sum_\mu a_\mu^2 b_{\mu,i}^2 
+ \sum_{\mu \neq \nu} 2 {\rm cos} [(\omega_\mu-\omega_\nu)t]
a_\mu a_\nu b_{\mu,i} b_{\nu,i}.
\eneq
The total energy
\beeq
\la \Psi | H | \Psi \ra = \sum_\mu a_\mu^2 \omega_\mu
\eneq
conserves, and this is checked, too.

Figure 3 shows the exact time development of probabilities
at twelve sites (the site numbers are indicated for D127 in
Fig. 1 (b)) of a wavefunction of the exciton.  Sample average 
over disorder has been also performed as before.  Probability 
amplitudes in Figs. 3 (a) and (b) are displayed in the 
logarithmic and linear scales, respectively.  Hereafter, 
time is measured in the units of $1/J$.  Series of plots are 
not labeled by the site numbers, but it is clear that 
all the series of plots for the sites 1, 2, ... and 12,
are ordered from the top to the bottom in Fig. 3 (a).  
The same symbols are used for the identical sites 1-12 
in Figs. 3-6.

The probability at the site 1 decreases within the time
of $t = 1 (1/J)$, and is nearly constant with the small
oscillation at longer time.  As shown clearly in Fig. 3 (b),
the decrease around $t=0$ is parabolic.  The coherence
of the exciton at the site 1 is lost within the time
scale $1 (1/J) = 1.4946$ fs.   In the course of time
development, the exciton tends to have more amplitudes 
at the site 2, 3, and so on.  However, the values of the
probability saturate finally, and small oscillations
continue in time afterward.  The saturated value at 
the site 1 is about 0.5, and it never becomes smaller 
than that of the site 2.  The probability at site 2 
is larger than that at the site 3.  Therefore, we 
find that the initial optical excitation captured at 
the outer edge of the dendrimer does not move into 
central regions of the molecule in the treatment of 
the equilibrium dynamics, although the time scale of 
1.4946 fs seems reasonable for intramolecular optical 
processes in organic systems.  Some kind of nonequilibrium 
effects would be necessary for exciton migrations to occur.

\subsection{Numerical time development}

Here, time is divided with a mesh of the width 
$\Delta t$, and time development of a wavefunction
of an exciton is calculated numerically.  The 
expansion $e^{-iH \Delta t} \simeq 1 - iH \Delta t 
+ O((\Delta t)^2)$ is retained up to the first order
of $\Delta t$.  The results are compared with those
of the exact dynamics.  We need such kind of the 
numerical treatment for nonequilibrium dynamics that 
will be studied in the next section.  Therefore, 
we perform numerical calculations of dynamics here
in order to compare with the exact results.

Figure 4 shows time developments of exciton propagation
calculated with the mesh $\Delta t = 0.0001 J$.  
Sample average over disorder has been done also.
Figures 4 (a) and (b) are for the short and long time 
dynamics, respectively.  The time variations of Fig. 4 (a)
are quite similar to those of Fig. 3 (a), even though
the absolute magnitudes of the saturated probabilities 
are somewhat different due to the linearized numerical
treatment.  In the long time behaviors show as Fig. 4 (b),
the probabilities of all the sites simply continue
oscillating.  The values of the probabilities tend to 
be larger when the site number is smaller.  Such the 
qualitative results remain the same as those of Fig. 3.  
In this sense, we can also use the numerical time 
developing treatment when the nonequilibrium effects 
are considered.  We note that energy of the system 
is kept constant while the simulation is continued 
within time of Fig. 3 (b).  Numerical errors do not 
seem developing as far as we look at the total energy.

\section{Nonequilibrium dynamics}

In this section, a positive damping factor $\Gamma$ 
is introduced as $J_{i,j} \rightarrow J_{i,j} - i \Gamma$ 
in the hamiltonian $H$, and time developments are 
investigated numerically with the time mesh $\Delta t = 
0.0001 J$.  In a toy model of an exciton dimer with the 
coupling $J_0 - i \Gamma$ ($J_0 > 0$), the two eigenvalues are
$\pm (J_0 - i \Gamma)$.  In the course of time developments,
there are two factors, $e^{-i J_0 t - \Gamma t}$ and
$e^{i J_0 t + \Gamma t}$.  They mean that the probability of
the wavefunction of the higher energy state decreases 
and the energy of the system moves to the lower energy state.
The introduction of a finite $\Gamma$ takes into account 
of damping of energy as in this toy model.  Such the 
nonequilibrium dynamics will generate motion of excitons 
from the outer edge to central areas in the dendrimers.  
The purpose of this section is to demonstrate the exciton 
flow in D127.

The similar method of including a damping term has been
used in the optical model of particle scattering to
the target nucleus [14].  The optical model has been
proposed in 1950's, and has been known in scattering problems.
The imaginary damping term is added to the hamiltonian.
And, this term describes well the effect that the 
incident particle is absorbed by the target with 
decreasing energy.

Development of a wavefunction of the exciton is
simulated numerically for the two damping factors:
$\Gamma = 0.1 J$ and $0.5J$.  The results are summarized 
in Figs. 5 and 6, respectively.  Figs. 5 (a) and 6 (a)
show behaviors in the short time scale $0 \leq t \leq
4 (1/J)$.  The decay of the exciton becomes a little bit 
faster than that of Fig. 4 (a), though the main trend 
remain similarly.  The dramatic differences are seen 
in the long time behaviors.  In Fig. 5 (b) for 
$\Gamma = 0.1J$, the probability at the site 1 becomes 
smaller than that of site 2 at the time larger than 
$40 (1/J)$.  The exciton at the outer edge begins motion 
in the direction of the molecular center due to the 
damping effects.  However, the probability at the
site 3 does not become larger than that of the site 1,
even though the probabilities at the sites with
large numbers are enhanced from those of $\Gamma = 0$
apparently.  Figure 6 (b) displays the case of the 
stronger damping $\Gamma = 0.5J$.   The probability
at the site 1 becomes smaller than that of the site 2
at time about $3 (1/J)$.  It even becomes smaller
than that of the site 12 (the center of the dendrimer)
at the time $\sim 40 (1/J)$.  Furthermore, the probabilities
of the site 2 and 3 become smaller than those of the
sites 4 and 5 which are located in the course of the 
optical path.  We thus find that {\sl the migration 
of an exciton really occurs owing to the damping effects}.

Figure 7 shows the variations of the total energy of
the system D127 as functions of time.  The plots indicate
the total energy after the mean energy $E$ is subtracted.
The squares, circles, and triangles are for $\Gamma = 0$ 
(Fig. 4), $0.1J$ (Fig. 5), and $0.5J$ (Fig. 6), respectively.  
The energy of the $\Gamma = 0$ case is constant as 
we described before.  However, in the cases of the 
finite $\Gamma$, the energy first decreases near $t=0$ 
and then increases, and become decreasing again.  
The time, where the energy begins decreasing secondly 
of the case $\Gamma = 0.5J$, is about $40 (1/J)$.  
This is near to the time when the two curves for the 
sites 1 and 12 cross in Fig. 6 (b).  The crossover
between the states located at the outer edge and the
states in the central area of the supermolecule {\sl really
occurs} in the time development.  Such the crossover to
the lower energy states is actually brought about by
the damping effects of the term $\Gamma$.  This 
theoretical fact demonstrates the necessity of the 
finite damping term in order that the dendrimer can 
act as a molecular antenna of energy from light.

\section{Summary}

We have proposed the coupled exciton model with off-diagonal 
disorder in order to characterize optical excitations in the 
extended dendrimers.  The decreasing optical excitation
energy with respect to the dendrimer size indicates
the presence of exciton pathways along the backbone
of the dendrimers.  The theoretical parameters are 
reasonable in view of the electron and hole hopping
processes, and also of the exciton decay time scale.

Next, dynamics of an exciton has been investigated 
by solving time development of a wavefunction
of optical excitations.   We have shown that a damping 
term with a certain magnitude is necessary in order 
that optical excitations captured at the outer edge 
of the supermolecule move to the central areas of 
the molecule.  It has been discussed that the damping
term gives rise to the crossover from states 
located at the outer edge to the states in the central 
area of the supermolecule in the time development.

\mbox{}

\begin{flushleft}
{\bf Acknowledgements}
\end{flushleft}

\noindent
Useful discussion with the members of Condensed Matter
Theory Group\\
(\verb+http://www.etl.go.jp/+\~{}\verb+theory/+),
Electrotechnical Laboratory is acknowledged.  
Numerical calculations have been performed on the DEC 
AlphaServer of Research Information Processing System 
Center (RIPS), Agency of Industrial Science and 
Technology (AIST), Japan.

\pagebreak
\begin{flushleft}
{\bf References}
\end{flushleft}

\noindent
$[1]$ Z. Xu and J. S. Moore, Acta Polym. {\bf 45},
83 (1994).\\
$[2]$ C. Devadoss, P. Bharathi, and J. S. Moore,
J. Am. Chem. Soc. {\bf 118}, 9635 (1996).\\
$[3]$ R. Kopelman, M. Shortreed, Z. Y. Shi, W. Tan,
Z. Xu, J. S. Moore, A. Bar-Haim, and J. Klafter,
Phys. Rev. Lett. {\bf 78}, 1239 (1997).\\
$[4]$ M. R. Shortreed, S. F. Swallen, Z. Y. Shi,
W. Tan, Z. Xu, C. Devadoss, J. S. Moore, and R. Kopelman,
J. Phys. Chem. {\bf 101}, 6318 (1997).\\
$[5]$ S. F. Swallen, Z. Y. Shi, W. Tan, Z. Xu,
J. S. Moore, and R. Kopelman, J. Lumin. {\bf 76\&77},
193 (1998).\\
$[6]$ R. W. Wagner, T. E. Johnson, and J. S. Lindsey,
J. Am. Chem. Soc. {\bf 118}, 11166 (1996).\\
$[7]$ A. Bar-Haim and J. S. Klafter, J. Lumin.
{\bf 76\&77}, 197 (1998).\\
$[8]$ K. Harigaya, Phys. Chem. Chem. Phys.
{\bf 1}, 1687 (1999).\\
$[9]$ T. Ritz, X. Hu, A. Damjanovi\'{c}, 
and K. Schulten, J. Lumin. {\bf 76\&77},
310 (1998).\\
$[10]$ F. C. Spano, J. R. Kuklinski, and S. Mukamel,
Phys. Rev. Lett. {\bf 65}, 211 (1990).\\
$[11]$ F. C. Spano, J. R. Kuklinski, and S. Mukamel,
J. Chem. Phys. {\bf 94}, 7534 (1991).\\
$[12]$ T. Kato, F. Sasaki, S. Abe, and S. Kobayashi,
Chem. Phys. {\bf 230}, 209 (1998).\\
$[13]$ ``Dendrimers", ed. F. V\"{o}gtle
(Springer, Berlin, 1998).\\
$[14]$ L. D. Landau and E. M. Lifshitz,
``Quantum Mechanics, 3rd ed." (Pergamon Press,
Oxford, 1977), p. 610.

\pagebreak

\noindent
TABLE I.  The energy of the lowest optical excitation.\\
\\
\begin{tabular}{lc} \hline \hline
Dendrimer &   Absorption edge\\ \hline
D4    & $E - 1.6037J$ \\
D10   & $E - 2.3346J$ \\
D25   & $E - 2.8078J$ \\
D58   & $E - 3.1588J$ \\
D127  & $E - 3.4338J$ \\ \hline \hline
\end{tabular}

\end{document}